\pdfoutput=1
\documentclass[preprint,aps,pra,a4paper,american,floatfix,pdftex,superscriptaddress,citeautoscript
]{revtex4-1}%
\usepackage{amsfonts,amsmath,amssymb}
\usepackage[T1]{fontenc}
\usepackage{graphicx}%
\usepackage{textcomp}
\usepackage[utf8]{inputenc}
\usepackage{microtype}
\usepackage{titlesec}
\usepackage{xspace}
\usepackage{xcolor}
\usepackage{hyperref, hypernat}
\usepackage{mathtools}
\usepackage{caption}
\usepackage{subcaption}

\newcommand*{\invcm}{\ensuremath{\text{cm}^{-1}}\xspace}%

\newcommand{\ruimm}{\affiliation{Radboud University, Institute for Molecules
      and Materials, Heyendaalseweg 135, 6525 AJ Nijmegen, The Netherlands}}%
\newcommand{\ucl}{\affiliation{Department of Chemistry, University College London, 20 Gordon Street, London, United Kingdom}}%

\begin{document}

\title{Preparation of tautomer-pure molecular beams by electrostatic deflection}

\author{Grite L. Abma} \ruimm
\author{Michael A. Parkes} \ucl
\author{\mbox{Daniel A.\ Horke}}%
\email[]{d.horke@science.ru.nl}%
\ruimm

\date{\today}%
\begin{abstract}
Tautomers are ubiquitous throughout chemistry, and typically considered inseparable in solution. Yet (bio)chemical activity is highly tautomer specific, with common examples being the amino and nucleic acids. While tautomers exist in an equilibrium in solution, in the cold environment of a molecular beam the barrier to tautomerization is typically much too high for interconversion, and tautomers can be considered separate species. Here we demonstrate the separation of tautomers and production of tautomerically-pure gas-phase samples. We show this for the 2-pyridone / 2-hydroxypyridine system, an important structural motif in both uracil and cytosine. Spatial separation of the tautomers is achieved via electrostatic deflection in strong inhomogeneous fields. We furthermore collect tautomer-resolved photoelectron spectra using femtosecond multiphoton ionization. This paves the way for studying the structure-function-dynamic relationship on the level of individual tautomers, using approaches that typically lack the resolution to do so, such as ultrafast dynamics experiments.
\end{abstract}
\maketitle

Tautomerization is ubiquitous across organic and biochemistry and commonly associated with any polar molecule containing acidic functional groups. Tautomers and their different tautomer-specific chemical functionalities play a key role in many biochemical processes. From enzyme catalysis~\cite{Vila:PNAS108:5602} and ligand binding~\cite{Thore:Science312:1208}, to DNA mutation and protein misfolding~\cite{sagvolden:JCPA114:6897,Taylor:PNAS63:253} and photo-stability of nucleic acids. While in solution the tautomer equilibrium can be heavily influenced by the solvent and solvent properties, most tautomeric systems are considered non-separable in solution. The situation in typical gas-phase experiments is very different. Here the extremely cold and collision-free environment of a supersonically expanded molecular beam usually prevents any tautomerization from occurring. However, since the transfer into the gas phase is so rapid, tautomer distributions typically remain nearly unchanged, with structures frozen into their respective local minima, rather than relaxing into the global minimum of the most stable tautomer. Hence gas-phase experiments still face the challenge of molecular samples containing multiple tautomers. While frequently these can subsequently be distinguished using high-resolution spectroscopy techniques, they can not be separated. This implies that properties such as relaxation dynamics can only be studied with tautomer resolution if the respective absorption wavelengths are sufficiently different for the tautomers. In the case of the 2-pyridone /2-hydroxypyridine tautomer system considered here, for example, the dynamics have been studied at specific excitation wavelength~\cite{min:chinjcp35:242,yang:pccp24:22710}, where it is possible to selectively excite or ionize only a single tautomer. Studies of the relaxation dynamics were both tautomers absorb, e.g. at wavelengths deeper in the UV such as 200~nm~\cite{hilal:MolSym45:165}, are not feasible with this approach. A case in point here are the still much debated electronic relaxation dynamics of cytosine, of crucial relevance for understanding DNA photostability. Different studies have reported vastly different excited-state lifetimes and differing numbers of relaxation  channels~\cite{Ullrich:PCCP6:2796,kosma:jacs131:16939,ho:jpca115:8406}. These discrepancies can, at least partly, be attributed to different tautomer contributions within the gas-phase samples~\cite{Ruckenbauer:scirep6:35522}. 

Similarly many modern gas-phase methodologies cannot distinguish different tautomer populations. In particular, experiments utilizing ultrafast laser pulses, such as attosecond charge-migration measurements~\cite{Calegari:Science346:336} or reaction pathway observations~\cite{Smith:PRL120:183003}, inherently lack the spectral resolution to distinguish tautomers. Moreover, molecular collision studies typically require pure samples of the collision partners~\cite{Chang:Science342:98,Tang:Science379:1031}, as do any imaging experiments based on electron or x-ray diffraction, which now have the potential for atomic temporal and spatial resolution~\cite{Barty:ARPC64:415,Kuepper:PRL112:083002,Hensley:PRL109:133202}. 

\begin{figure}
    \includegraphics[width=0.5\textwidth]{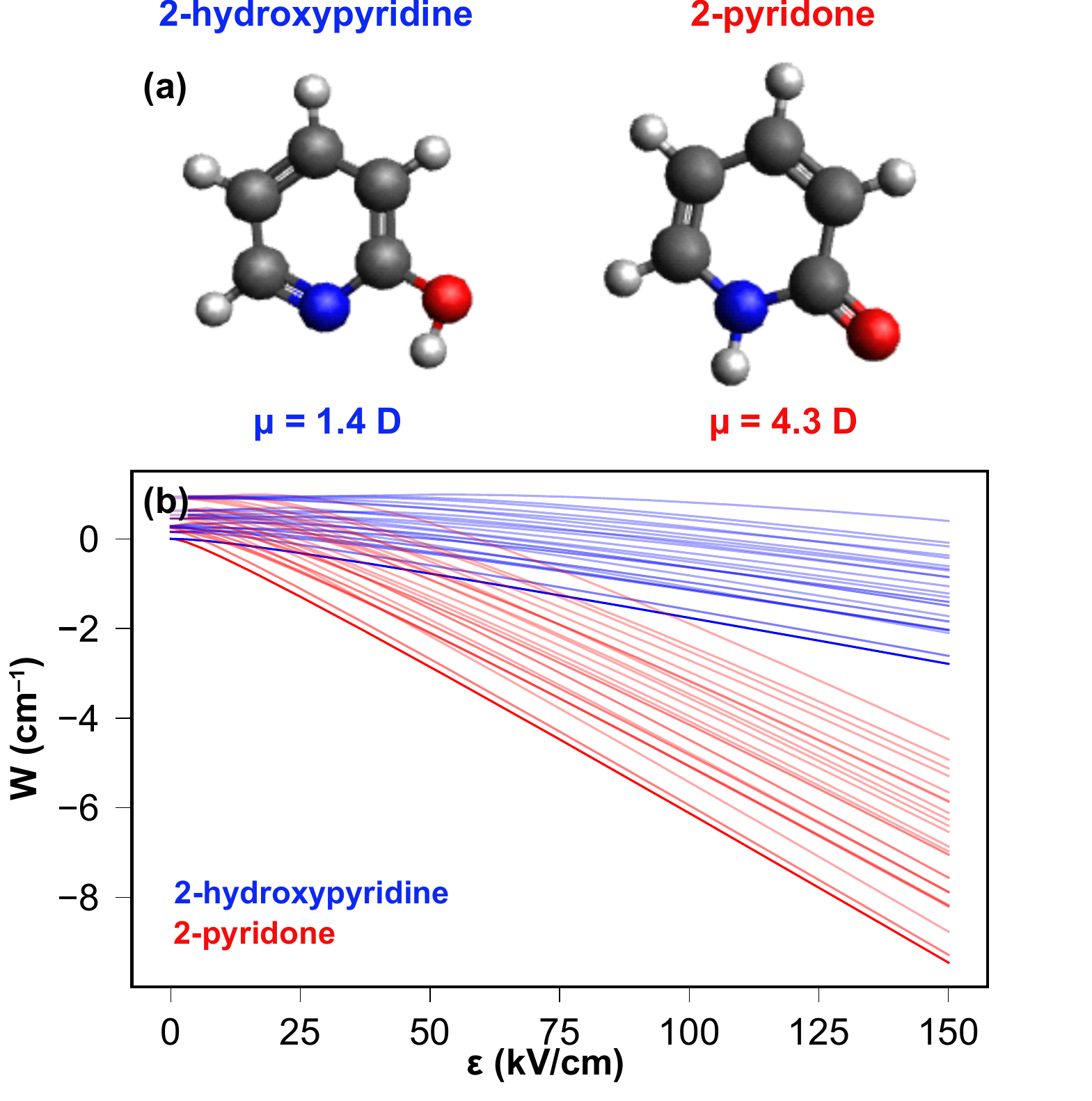}
    \caption{Structures and dipole moments of 2-hydroxypyridine and 2-pyridone (a), and the associated Stark shift as a function of the electrostatic field (b). These are shown for rotational states $J=0-2$ and clearly show the different extent of the Stark interaction for the two tautomers.}
    \label{fig:intro}
\end{figure}

We show here spatial separation of tautomers in the gas-phase with the production of a tautomerically-pure cold molecular beam, by utilizing the established electrostatic deflection technique~\cite{Holmegaard:PRL102:023001,Filsinger:ACIE48:6900,Chang:IRPC34:557}. In particular we demonstrate this approach for the 2-hydroxypyridine (\emph{enol} or \emph{lactim} tautomer)/2-pyridone (\emph{keto} or \emph{lactam} tautomer) system (\autoref{fig:intro}), a structural motif for both uracil and cytosine and a common model system for DNA tautomerization. Both tautomers are stable in the gas phase, with a barrier of 34-38~kcal/mol between them. The 2-hydroxypyridine tautomer is slightly lower in energy by about 300~\invcm, which leads to an expected tautomer distribution of about 3:1 (2-hydroxypyridine:2-pyridone)~\cite{yang:pccp24:22710} for pure samples at room temperature, whereas in polar solvents and the crystal phase 2-pyridone dominates~\cite{nowak:jcp9:1562}. Additionally, this system is known to form very stable dimers in the gas phase due to strong N--H and O--H hydrogen bonding interactions, making it an ideal model system for DNA hydrogen bonding.

In order to control and separate the two tautomers we use the electrostatic deflection approach~\cite{Holmegaard:PRL102:023001,Filsinger:ACIE48:6900,Chang:IRPC34:557}. This exploits the difference in the permanent dipole moments of the tautomers, leading to a species-specific Stark interaction with an electrostatic field~\cite{Chang:IRPC34:557}, as shown in \autoref{fig:intro}(b). This shows the calculated Stark energy shift for the rotational levels $J=0-2$ for both tautomers, calculated using the freely available CMIstark software package~\cite{Chang:CPC185:339}. It clearly highlights that, due to their distinct permanent electric dipole moments, both tautomers exhibit significantly different Stark shifts. We utilize this by passing the tautomers through an electrostatic deflector with a strong inhomogeneous electric field inside~\cite{Holmegaard:PRL102:023001}, leading to the exertion of a tautomer-specific force and hence acceleration. This transverse acceleration then leads to a spatial separation of species based on their effective dipole-moment to mass ratio~\cite{Chang:IRPC34:557}. This approach has previously been successfully demonstrated for the separation of, amongst others, individual quantum states of small (triatomic) molecules~\cite{Nielsen:PCCP13:18971,Horke:ACIE53:11965}, rotational isomers~\cite{Filsinger:PRL100:133003,Filsinger:ACIE48:6900,Teschmit:ACIE57:13775} and clusters~\cite{Trippel:PRA86:033202, Bieker:JPCA123:7486, Johny:CPL721:149}. Here we use this approach to separate the two tautomers. We moreover collect photoelectron spectra of the tautomer-separated samples, allowing us to extract and assign tautomer-specific photoelectron spectra using femtosecond multiphoton ionization (fs-MPI). Our separation measurements are supported by trajectory simulations and all spectral assignments are confirmed by quantum-chemical calculations.

\begin{figure}
    \includegraphics[width=0.5\textwidth]{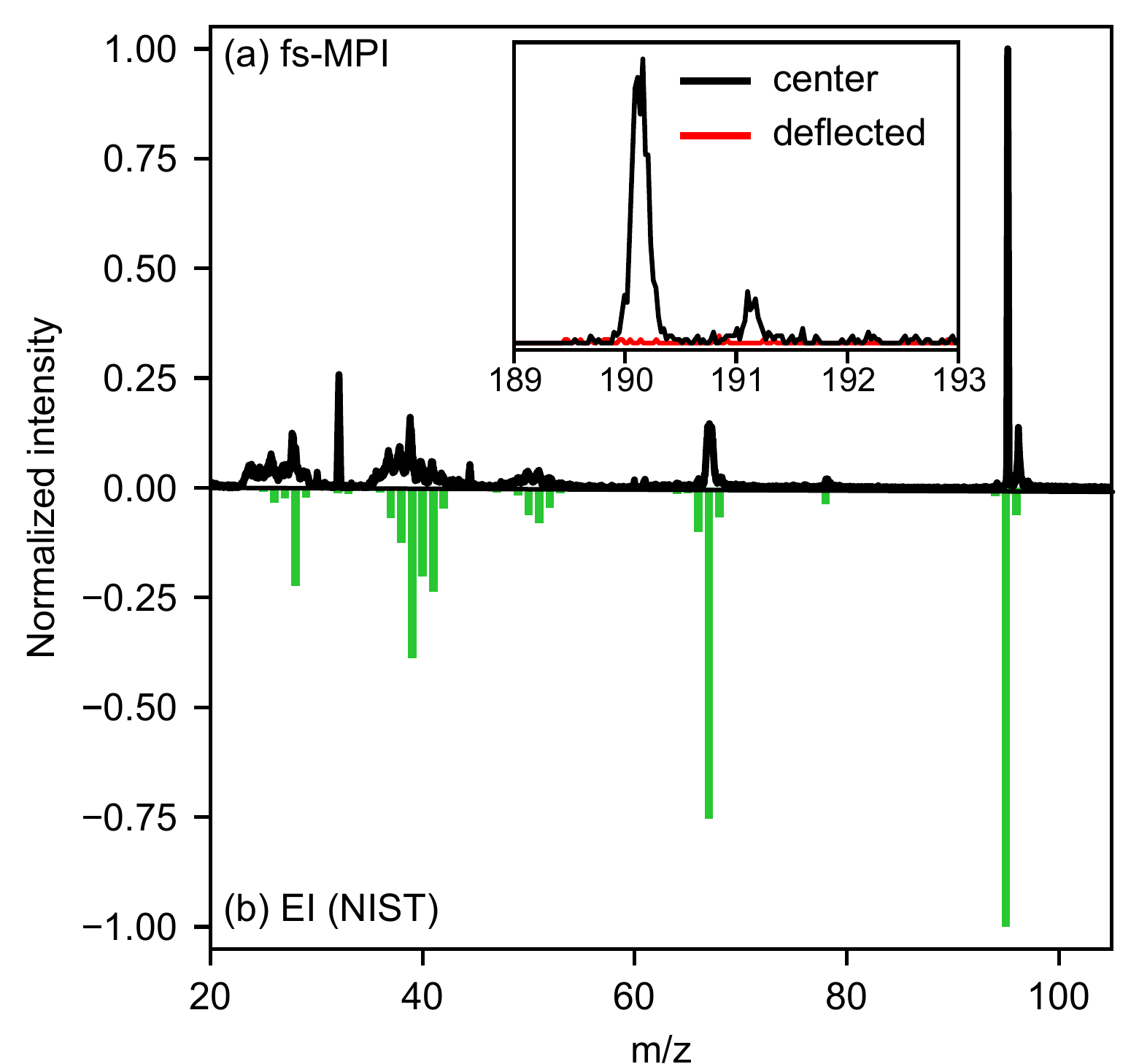}
    \caption{Mass spectrum of 2-hydroxypyridine/2-pyridone collected with 400~nm fs-MPI (a), and a reference spectrum from NIST collected with electron impact ionization (b)~\cite{NIST:webbook:MassSpec}. The inset shows an enlarged view of the mass region containing dimer signal. In the center of the molecular beam (black trace) the dimer signal amounts to around 5\% of the monomer signal, whereas in the deflected beam (red trace) no dimers are observed, see text for further details.}
    \label{fig:massspec}
\end{figure}
Typical mass spectra measured in the molecular beam following fs-MPI are shown in \autoref{fig:massspec}(a). These are dominated by the parent ion at m/z=95, although some fragmentation is observed with significant channels at m/z=28, 39 and 67. In panel (b) we show a reference mass spectrum from NIST obtained using electron impact ionization~\cite{NIST:webbook:MassSpec}, which exhibits identical fragmentation channels, with an overall larger degree of fragmentation. This confirms our molecular assignment and highlights the relative softness of the fs-MPI process~\cite{Dauletyarov:JASMS34:1538}. Pyridone is known to form very stable dimers, and these were also present in our mass spectrum. Even under supersonic expansion conditions optimized to minimize dimer formation, we observed a signal in the dimer channel with an intensity of around 5\% that of the monomer, as shown in the inset of \autoref{fig:massspec}. 2-pyridone can form three different dimers, a homomeric dimer consisting of two 2-hydroxypyridine units, a homomeric dimer of two 2-pyridone units and a mixed dimer. At a temperature of 400~K, the expected relative populations of the dimers are 1.3\%, 98.4\% and 0.4\% for the (2-hyroxypyridine)$_2$ (\emph{enol-enol}), (2-pyridone)$_2$ (\emph{keto-keto}) and mixed (\emph{keto-enol}) dimers~\cite{slanina:jms257:491}, respectively. Hence the 2-pyridone dimers are expected to dominate in the molecular beam, as experimentally shown by Held et al~\cite{held:JACS112:8629}. This also follows from computational optimization of the dimer structure, as the mixed dimer optimizes to the structure of the (2-pyridone)$_2$ dimers through hydrogen shuttling, indicating that this dimer is the lowest energy configuration. However, other studies have previously observed both dimer systems~\cite{muller:jcp115:5192,borst:chemphys283:341}, without quoting relative abundances. The dimers consisting of a single tautomer do not exhibit a dipole moment due to their inherent symmetry. The mixed dimer is expected to have a dipole moment of about 5~D, however, as outlined above, it is not expected to have any significant population in the molecular beam.

\begin{figure}
    \includegraphics{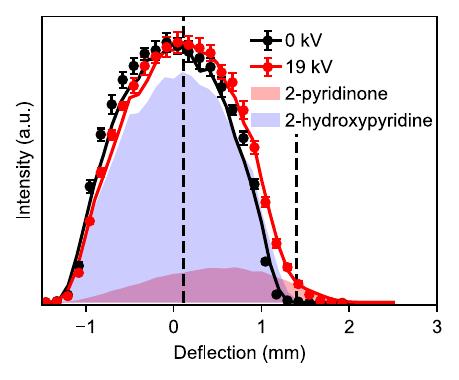}
    \caption{Spatial density profiles of the molecular beam without electrostatic field (black) and with electrostatic field present (red). Measurements are shown as data points with error bars, while solid lines correspond to trajectory simulations. The shaded areas correspond to the tautomer-resolved densities extracted from the simulations. These show that for the deflected beam, at positions $>1.2$~mm a pure sample of 2-pyridone is created.}
    \label{fig:deflection}
\end{figure}

In order to separate the tautomers, the molecular beam was passed through an electrostatic deflector, with a potential difference of 19~kV applied across the electrodes (see S.I. for details). The resulting deflection of the molecular beam was monitored by translating the tightly focused probe laser beam and recording mass spectra at various positions, yielding a spatial density profile. This is shown in \autoref{fig:deflection} for the m/z=95 channel for both the field-free case (0~kV, black data points) and with the strong field present (19~kV, red data points). In the field-free case a symmetric distribution centered at 0~mm was observed, which in the presence of the strong field deflected to the right and developed a noticeable edge on the more deflected side, around 1.5~mm.

Since this data is extracted from fs-MPI and time-of-flight mass spectrometry, it is not tautomer-specific. In order to understand the relative tautomer distributions within the beam, we have performed trajectory simulations for both tautomers under realistic experimental conditions~\cite{Chang:IRPC34:557}. These simulations were then fitted to the experimental data, yielding an initial rotational temperature of 3.4~K and a 2-pyridone fraction of 16\% in the molecular beam, as discussed in the supplementary information. Simulated spatial density profiles for the individual tautomers are shown as shaded areas in \autoref{fig:deflection}, whilst the sum of the two contributions is shown as a solid line. The simulation results are in excellent agreement with the experimental data points, and show that at spatial positions $>1.2$~mm a pure beam of 2-pyridone is produced. We note that the deflected beam is also free from contamination by the dimer system, as expected based on the known dimer distributions and dipole moments. This is confirmed by the mass spectrum around m/z=190, as shown in the inset of \autoref{fig:massspec}, with no detectable dimer signal in the deflected beam (see SI for further details). 

In order to further confirm the tautomer separation in the deflected beam, we measured photoelectron images in the center of the beam and in the deflected edge, at the positions indicated by the dashed lines in \autoref{fig:deflection}, which correspond to 0.1 and 1.4~mm above the center of the molecular beam. These measurements were again conducted using fs-MPI. The ionization energies for 2-pyridone and 2-hydroxypyridine have previously been determined as 8.93~eV and 8.45~eV, respectively~\cite{Ozeki:JPC99:8608}. Thus necessitating 3 photons at 400~nm (total photon energy 9.3~eV) for ionization. 
\begin{figure}[t!]
    \includegraphics[width=0.5\textwidth]{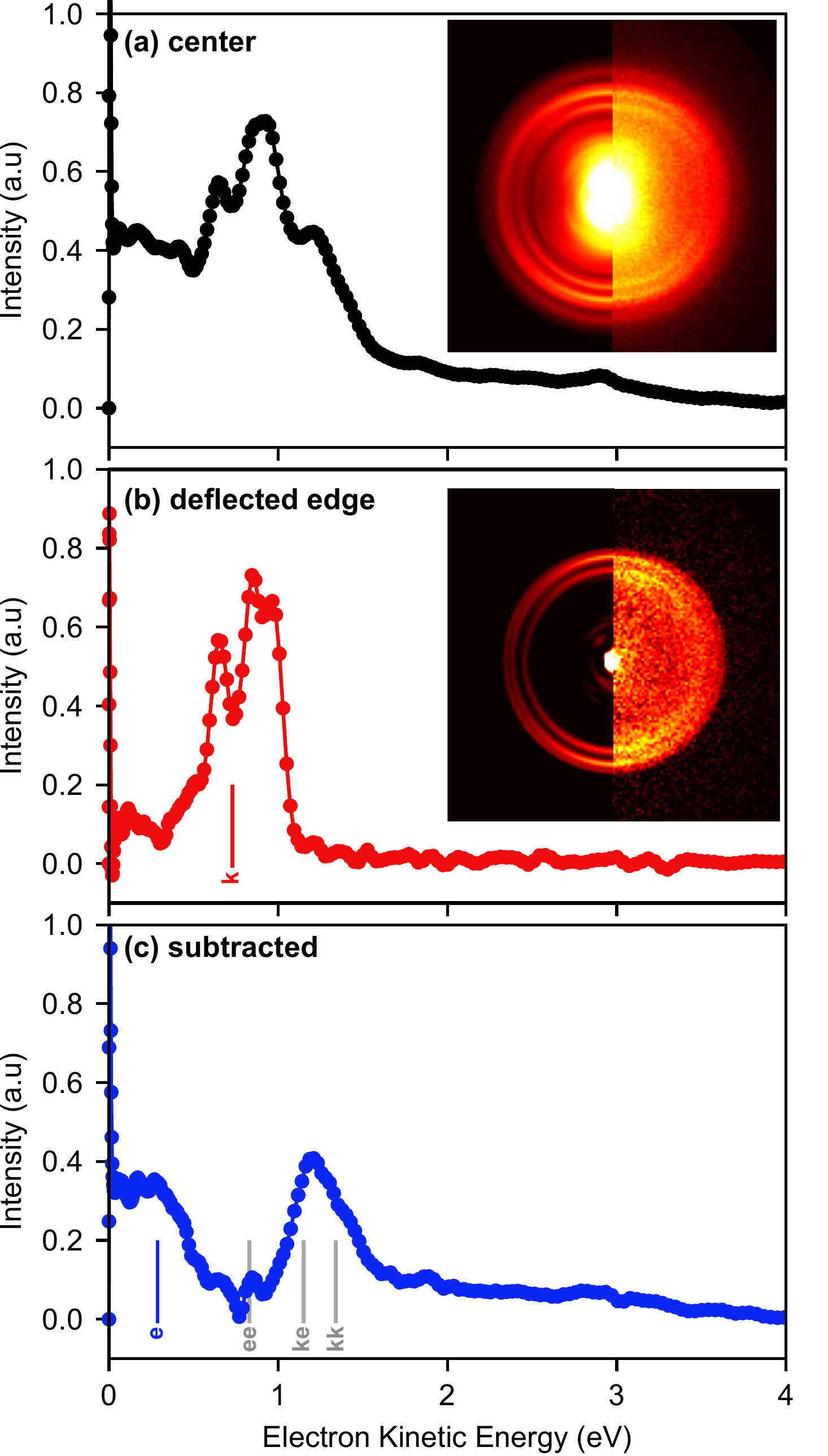}
    \caption{Photoelectron images and extracted photoelectron spectra in (a) the center of the beam and (b) the deflected edge of the molecular beam. The latter corresponds to a spectrum of pure 2-pyridone. (c) Mixed photoelectron spectrum with the contribution from the 2-pyridone subtracted. This contains features from both 2-hydroxypyridine, as well as from dimers. Calculated vertical ionization energies are shown as solid bars throughout. For the dimer system calculated energies for all three stoichiometries (\emph{enol-enol}(ee), \emph{keto-enol} (ke) and \emph{keto-keto}(kk)) are shown.}
    \label{fig:pes}
\end{figure}

Collected photoelectron images and spectra are shown in \autoref{fig:pes}. We note that all images show a sharp signal in the center of the image, corresponding to 0~eV electrons, that originate from residual water. In panel (a) we show the data collected at 0.1~mm above the center of the molecular beam, and therefore containing electrons originating from both tautomers, as well as contributions from the dimer systems. The photoelectron spectrum was extracted using Abel inversion~\cite{Hickstein:pyabel,rBasex:0.9.0}, and shows several distinct peaks in particular in the range 0--2~eV of electron kinetic energy. The corresponding image and photoelectron spectrum for the deflected beam are shown in \autoref{fig:pes}(b), and show only a subset of the features present in (a). We assign this spectrum collected in the deflected beam to the 2-pyridone \emph{keto} tautomer. The known ionization energy should yield a maximum electron kinetic energy of 0.9~eV, agreeing well with our data when taking into account the bandwidth of a 3-photon fs-MPI process. We have furthermore calculated the expected vertical ionization energy for the formation of the $X^{0}_{0}$ cation ground state, which yielded an electron kinetic energy of 0.73~eV (indicated by the vertical bar in \autoref{fig:pes}(b)), in good agreement with the experimental data. The experimental spectrum contains 3 closely spaced features approximately  0.11~eV (890~\invcm) and 0.20~eV (1610~\invcm) apart, most likely arising from vibrational excitation of in-plane bending modes upon ionization~\cite{Frey:JCP125:114308}.

In \autoref{fig:pes}(c) we show a photoelectron spectrum of the mixed beam (a) with the pure-\emph{keto} spectrum (b) subtracted after normalization on the 2-pyridone signal. The resulting difference spectrum contains two major features. We assign the low energy feature to formation of the $X^{0}_{0}$ cation ground state in the 2-hydroxypyridine tautomer. The calculated vertical ionization energy for this yields electrons with kinetic energies of 0.23~eV (blue vertical bar in \autoref{fig:pes}(c)), whilst the published adiabatic ionization energy corresponds to an energetic cutoff at 0.4~eV, both in agreement with our data. This energetic cut-off also implies that the high energy feature observed cannot correspond to the \emph{enol} tautomer. Instead we attribute this feature to the dimers present in the molecular beam. We have calculated the vertical ionization energies for the formation of the cation ground state for all the three potential dimer systems (yielding expect electron kinetic energies of \emph{enol-enol}, 0.83~eV; \emph{keto-enol}, 1.15~eV and \emph{keto-keto}, 1.34~eV ), as indicated by the grey vertical bars in \autoref{fig:pes}(c). The major peak observed energetically agrees with the calculated vertical ionization energies of both the \emph{keto-enol} and \emph{keto-keto} geometries. However, given that we did not observe any deflection in the dimer channel, we assign this peak to the \emph{keto-keto} dimer structure, in agreement with previous experimental studies~\cite{held:JACS112:8629} and the relative dimer stabilities, that predict $>98$\% of dimers to be in the \emph{keto-keto} geometry~\cite{slanina:jms257:491}. From the collected spectrum we obtain values of the vertical and adiabatic ionization energy of the dimer of 8.1~eV and 7.7~eV, respectively. Both the feature attributed to the \emph{enol} tautomer, as well as that due to the dimer, are absent from the photoelectron spectrum recorded in the deflected molecular beam (\autoref{fig:pes}(b)), confirming the creation of a pure tautomer-selected sample.

In summary, we demonstrated here the separation of tautomers in the gas phase, utilizing electrostatic deflection, for the important 2-pyridone/2-hydroxypyridine system. This enabled the creation of a pure sample of the \emph{keto} tautomer, which furthermore was free from contamination by the strongly-bound dimers.  Experimental results showed excellent agreement with detailed trajectory simulations, and were furthermore corroborated via photoelectron spectroscopy with a universal 400~nm femtosecond multiphoton ionization process. Recorded spectra could be assigned using quantum chemical calculations. The methodology used here is widely applicable and only requires that the molecule of interest can be entrained in a cold molecular beam and that the tautomers have a sufficient difference in dipole moment to mass ratio~\cite{Chang:IRPC34:557}. This makes this approach highly applicable to many important biological building blocks, such as cytosine~\cite{Franz:epjd68:279}. The spatial dispersion of tautomers can generally add tautomer sensitivity to experimental techniques that otherwise inherently lack the resolution to distinguish them, as we have shown here for femtosecond multiphoton ionization. This paves the way for ultrafast dynamics experiments or time-resolved imaging studies focusing on the biologically relevant tautomers of important biomolecules.

\section*{Experimental Methods}
All experiments were performed in a custom-build molecular beam setup, equipped with an Even-Lavie valve~\cite{Even:AIC2014:636042}, 30~cm long electrostatic deflector~\cite{Holmegaard:PRL102:023001,Filsinger:ACIE48:6900,Chang:IRPC34:557} and velocity-map imaging (VMI) spectrometer~\cite{Eppink:RSI68:3477}, further technical details are given in the supporting information. Chemical samples were commercially available (97\% purity, Sigma-Aldrich) and used without further purification. Molecules were ionized via femtosecond multiphoton ionization (fs-MPI) using 400~nm laser pulses of 100~fs duration.

Experimental deflection measurements were complemented by trajectory simulations. For these the Stark effect of both isomers was calculated using the freely available CMIstark software package~\cite{Chang:CPC185:339}. Trajectories of individual quantum states ($J=0-15$, 2000 trajectories per state) were then propagated through the experimental setup~\cite{Chang:IRPC34:557}. Individual trajectories were combined to a simulated deflection profile through Boltzmann weighting, with the rotational temperature and the tautomer distribution fitted to the experimental data.

The structures of 2-pyridone, 2-hydroxypyridine and all possible dimers were optimized using M\"oller Plesset perturbation (MP2) theory with a 6-311G++(3df, 3pd) basis set in Gaussian16~\cite{Gaussian:2016}. The structures were checked to be minima by performing frequency calculations. These optimized structures were then used as inputs for EOM-IP-CCSD calculations in Q-Chem 5,~\cite{qchem5} with a 6-311G** basis set to calculate vertical ionization energies.

\clearpage
\section*{Acknowledgements}
This work was supported by the Netherlands Organization for Scientific Research (NWO) under grant numbers STU.019.009, 712.018.004 and VI-VIDI-193.037 and by the European Research Council (ERC) under the European Union’s Horizon 2020 Research and Innovation Program (grant Agreement no. 817947 FICOMOL). We would like to thank Prof. Rick Bethlem for the loan of the electrostatic deflector. We furthermore thank the Spectroscopy of Cold Molecules Department, and in particular Prof. Bas van de Meerakker, for continued support. Computational parts of this work were carried out on the Dutch national e-infrastructure with the support of SURF Cooperative.

\paragraph*{Supporting information avaialable:} Technical details of molecular beam apparatus, details of trajectory simulations, deflection profiles for dimer, optimized geometries.

\clearpage
\bibliography{string,UCD-bib}
\bibliographystyle{achemso}

\end{document}